 \documentstyle[12pt]{article}
 \newcommand{\be}{\begin{equation}}
 \newcommand{\ee}{\end{equation}}
 \newcommand{\ba}{\begin{eqnarray}}
 \newcommand{\ea}{\end{eqnarray}}
 \makeatletter
 \@addtoreset{equation}{section}
 \makeatother
 \renewcommand{\theequation}{\thesection.\arabic{equation}}
 \def\appendix #1
 {
  \section*{Appendix: #1}
  \renewcommand{\theequation}{A.\arabic{equation}}
 }

 \def\op#1{\mathop{\fam0 #1}\limits}
 \def\Spin{\hbox{\rm Spin}}
 \def\SO{\hbox{\rm SO}}
 \def\GL{\hbox{\rm GL}}
 \def\Met{\hbox{\rm Met}}
\def\sinh{\hbox{\rm sinh}}
\def\cosh{\hbox{\rm cosh}}

 \def\Im{\hbox{\rm Im}}

 \def\Aut{\hbox{\rm Aut}}

 \def\calC{{\cal C}}
 \def\calL{{\cal L}}
 \def\tarr{\leadsto}
 \def\d{\partial}

 \def\La{\Lambda}
 \def\la{\lambda}
 \def\Si{\Sigma}
 \def\sig{\sigma}

 \def\ep{\epsilon}
 \def\al{\alpha}
 
 \def\Ga{\Gamma}
 \def\ga{\gamma}
 \def\De{\Delta}

 \def\vp{\varphi}
\def\na{\nabla}

\def\Re{I \kern-.36em R}              
\def\Co{I \kern-.66em C}              
\def\del{\partial}                    

 \def\arr{\rightarrow }                
 \def\rarr{\longrightarrow }           
 \def\larr{\longleftarrow }            
 
 \mathchardef\lhook="312C
 \mathchardef\rhook="312D
 \def\lharr{\larr\joinrel\rhook}
 \def\rharr{\lhook\joinrel\rarr}
 \def\PRD#1#2#3{{\it Phys.\ Rev.} {\bf D#1}, #2 (#3)}
 \def\PRL#1#2#3{{\it Phys.\ Rev.} Lett.\ {\bf #1}, #2 (#3)}

 \def\NPBPS#1#2#3{{\it Nucl.\ Phys.\ (Proc.\ Suppl.)} {\bf B#1}, #2 (#3)}
 
 \def\PLB#1#2#3{{\it Phys.\ Lett.} {\bf B#1}, #2 (#3)}

 \def\JMP#1#2#3{{\it J.\ Math.\ Phys.} {\bf #1}, #2 (#3)}

 \def\IJMPA#1#2#3{{\it Int.\ J.\ Mod.\ Phys.} {\bf A#1}, #2 (#3)}
 
 \def\MPLA#1#2#3{{\it Mod.\ Phys.\ Lett.} {\bf A#1}, #2 (#3)}
 
 \def\CQG#1#2#3{{\it Class.\ Quantum Grav.} {\bf #1}, #2 (#3)}
 
 \def\ANP#1#2#3{{\it Ann.\ Physics (N.Y.)} {\bf #1}, #2 (#3)}

\begin{document}
\title{Two-dimensional dilaton gravity coupled to massless spinors}
\author{Marco Cavagli\`a,\thanks{E-mail:
cavaglia@aei-potsdam.mpg.de}\\
\small\sl Max-Planck-Institut f\"ur Gravitationsphysik,
\small\sl Albert-Einstein-Institut\\
\small\sl Schlaatzweg 1, D-14473 Potsdam, Germany\\
\\
Lorenzo Fatibene\thanks{E-mail: fatibene@dm.unito.it}\\
\small\sl Dipartimento di matematica,
\small\sl Universit\`a di Torino\\
\small\sl Via Carlo Alberto 10, I-10123 Turin, Italy\\
\small and\\
Mauro Francaviglia\thanks{E-mail: francaviglia@dm.unito.it}\\
\small\sl Dipartimento di matematica,
\small\sl Universit\`a di Torino\\
\small\sl Via Carlo Alberto 10, I-10123 Turin, Italy\\
}

\date{\today }
\maketitle

\begin{abstract}
We derive exact solutions of two-dimensional dilaton gravity coupled to
massless spinors for some particular choices of the dilatonic potential.
For constant dilatonic potential the model turns out to be completely
solvable and the general solution is found. For linear and exponential
dilatonic potentials we present the class of exact solutions with a
Killing vector. 

\medskip
\noindent
%
Pac(s) numbers: 04.20.Gz, 04.20.Jb, 04.60.Kz\\
Keyword(s): Spinor structures, Exact solutions, Reduced Models.
\end{abstract}
\section{Introduction\label{intr}}
Recently, two-dimensional dilaton-matter-gravity (DMG) theories have
attracted much attention because of their connection to string theory and
also, for some choices of the dilatonic potential and matter fields, to
dimensionally reduced models of gravity in $N\ge 3$ dimensions (see, for
instance, \cite{strom}-\cite{cava} and references quoted therein.)
Furthermore, DMG models can be seen as simplified models which are useful to 
clarify some conceptual problems of quantum gravity. So several 1+1 models of 
DMG have been discussed in the literature, both from the classical and quantum
points of view \cite{cava}-\cite{2d5}. In this context, attention
has been focused essentially on dilaton-gravity (DG) models coupled to
gauge and scalar fields. (For short reviews, see e.g.\ 
\cite{jacksm,filippov}.) Not surprisingly, we are not aware of papers
dealing with models of two-dimensional DG coupled to fermionic fields. The
lack of discussion about two-dimensional spinor-dilaton-gravity (SDG)
models is indeed somehow related to the absence of a general formalism
able to describe gravity dynamically coupled to spinorial fields. The aim
of this paper is to fill this gap partially and present some exact
solutions of two dimensional SDG. We shall use a general formalism for
spinor-gravity theories that is well posed from a geometrical point of
view and is potentially able to deal with interactions between gravity and
spinors without any restriction on dimension or signature (see
\cite{VB,Spinors}). The formalism also describes a truly relativistic
field theory, in the sense that it does not make use of any background
field (such as fixed background metrics) in contrast to what is often done
when dealing with spinors. As a consequence, we are able to describe both
the effect of the gravitational field on spinors (as in the case of
`background theories' on curved spaces) as well as the effect of spinors
on the gravitational field itself. 

The framework recalls gauge theories in their geometrical formulation
where one starts from a principal bundle on spacetime $M$, the so-called
structure bundle $\Si$, that encodes the symmetry structure of the
theory. The configuration bundle $\calC$ is then associated to the
structure bundle, i.e.\ the principal automorphisms of the structure
bundle are represented on the configuration bundle by means of a canonical
action. Automorphisms of the structure bundle are requested to be
symmetries of the Lagrangian $L$, namely to act on the configuration
bundle leaving the scalar Lagrangian unchanged. 

Gravity is described by new variables, called spin frames, suitably
related to spin structures on spacetime and defined without any reference
to any preferred background metric. So at a fundamental level no metric
appears at all. The metric is canonically induced by the spin frame
(see equation (\ref{ind-met})) once the field equations have been solved: 
\medskip
\be
(M,\Si)  \tarr (\calC,L)  \tarr
\begin{array}{l}
\hbox{field equations}\\
\hbox{dilaton+Einstein+}\\
\hbox{+Dirac equations}\\
\end{array}
\tarr
\begin{array}{l}
\hbox{spinors}\\
\hbox{spin frames}
\tarr
\hbox{a geometry $g$ on $M$} \\
\hbox{other fields.}\\
\end{array}
\ee
\medskip
Locally, spin frames resemble vielbein because both assign a $\GL(N)$
matrix to each point of spacetime $M$ ($\dim(M)=N$). However, spin frames
and vielbein transform differently. Vielbein are natural objects,
i.e.\ they transform naturally under both diffeomorphisms of $M$ and the
appropriate representation of the orthogonal group, and these two actions
are completely unrelated; spin frames transform just under automorphisms
of the structure bundle $\Si$ (see equation (\ref{Spin-frames-transf}) in the
appendix), as a kind of gauge fields. 

Let us mention here that the spin frame formalism allows to overrun an
obstacle that is usually encountered in the discussion of spinor-gravity
theories. In the standard approach one needs vielbein to write the Dirac
Lagrangian because the world indices of the covariant derivatives
must be transformed into vielbein indices to be contracted with
Dirac matrices.  This prevents the theory from being geometrically well defined
since in this case vielbein are dynamical fields and have to be varied to
obtain the field equations. However, vielbein are intrinsically local
objects, i.e.\ they are local sections of the frame bundle $L(M)$. On
reasonably general manifolds, i.e.\ non-parallelizable ones, there is no
global section of $L(M)$ at all. 

In this paper we are just interested in field equations and exact
solutions so local variational techniques provide (local) field equations
which are invariant under gauge transformations. Due to this invariance,
local solutions fit together smoothly on patches and we can always glue
them together to obtain a global solution. However, we prefer to
introduce the spin frame formalism in view of forthcoming investigations
such as, for instance, the study of conserved quantities (that are not local
quantities) and higher-dimensional models. In this perspective, the
standard formalism is not adequate since one needs a global frame to
define a global Lagrangian and spinor theories make sense just on
parallelizable manifolds. Unlike ordinary frames, spin frames are
geometrically well defined in any dimension and for any signature. 
Moreover, they are guaranteed to be continuous and globally defined on
every spin manifold, i.e.\ on a much wider class of manifolds than the
class of parallelizable ones. (Of course, in two dimensions any spin
manifold is parallelizable so, for the purposes of this paper, we do not
really need spin frames.)

Since here we simply derive exact solutions, we relegate to the appendix
the discussion of the spin frame formalism and technical details.
\section{Action and field equations\label{Action}}
From now on, we shall work locally. The global framework (see the appendix) 
may be recovered at the end, when both the manifold structure of $M$ and
the structure bundle $\Si$ over it can be obtained by gluing local
solutions.

Let us consider massless spinors. (The massive case will be discussed
elsewhere.)  The SDG Lagrangian density is (we follow the conventions of
\cite{weinberg})
\be
\calL=\sqrt{-g}\left[-\phi R+U(\phi)+i\alpha\left(
\bar \psi \gamma^a \na_a\psi-\na_a\bar \psi\gamma^a
\psi \right)\right]\,,
\label{Lag}
\ee
where $R$ is the scalar curvature of the metric $g_{\mu\nu}$ induced by a
spin frame $\La$ on $\Si$, $\sqrt{-g}$ is the square root of (the absolute
value) of its determinant, $\gamma^a$ are the two-dimensional Dirac
matrices defined in the appendix, and $\alpha$ is a (real) coupling
constant. We have denoted the conjugate spinors by $\bar
\psi=\psi^\dagger\cdot\ga_0$ and the covariant derivative of spinors by
\be
\na_a \psi=e^\mu_a\left(d_\mu \psi+{1\over8}[\gamma_b,\gamma_c]\psi
\Gamma^{bc}_\mu\right)\,,
\qquad
\na_a \bar \psi=e^\mu_a\left(d_\mu \bar \psi-{1\over8}\bar \psi
[\gamma_b,\gamma_c]\Gamma^{bc}_\mu\right)\,,
\ee
where $\Ga^{ab}_\mu=e^a_\rho(\Ga^\rho_{\sig\mu}e^{b\sig}+d_\mu
e^{b\rho})$ are the coefficients of the spin connection, and
$\Ga^\rho_{\sig\mu}$ are the Christoffel symbols of the induced metric
$g_{\mu\nu}$. The function $U(\phi)$ is the dilatonic potential.

Equation (\ref{Lag}) is the local expression of a global $\Aut(\Si)$-covariant
Lagrangian and defines the action
\be
S_D=\int_D \calL\>d^2x\,,
\label{Act}
\ee
where $d^2x$ is the local volume element and $D\subset M$ is a compact
two-dimensional submanifold with a compact one-dimensional boundary $\del
D$. Varying the action (\ref{Act}) we obtain the field equations
\ba
&&\gamma^a\na_a \psi=0\,,\label{EDirac}\\
&&-R+{d U\over d\phi}=0\,,\label{EDilatone}\\
&&(\nabla_{(\mu}\nabla_{\nu)}-g_{\mu\nu}\nabla_\sigma\nabla^\sigma)\phi+{1\over
2}Ug_{\mu\nu}=H_{\mu\nu}\,,\label{EEinstein}
\ea
where $H_{\mu\nu}$ is the Hilbert stress tensor evaluated on-shell,
i.e.\ modulo terms that vanish on solutions:
\ba
H_{\mu\nu}={i\alpha\over 2}
\Big(\na_{a}\bar \psi\gamma_{b} \psi-\bar \psi\gamma_{b}
\na_{a}\psi\Big) e^a_{(\mu} e^b_{\nu)}\,.\label{hilbert}
\ea
Since the Lagrangian density $\calL$ is covariant with respect to
automorphisms of $\Si$, these transformations map solutions into
solutions, leaving the field equations invariant. If $(e_a^\mu, \phi,
\psi)$ is a solution of the field equations over the open set $U_\al$,
then
\be
(e_a^\mu, \phi,\psi)
\quad\leadsto\quad
(J^\nu_\mu\>e_b^\nu\>\ell^b_a(\vp^{-1}), \phi, \vp\cdot \psi)
\label{OutOfGauge}
\ee
is also a solution over the open set $f(U_\al)$ whatever automorphism
$\Phi:\Si\arr\Si$ projecting onto the diffeomorphism $f:M\arr M$ is
considered.

\section{Exact solutions\label{sol}}
Since the manifold $M$ is two-dimensional any metric $g$ can be locally
cast into the conformal form
\be
g=4\rho^2(u,v) \cdot { {1\over  2}}(du\otimes dv+ dv\otimes
du)\,,\label{ansatz-metric}
\ee
where $u$ and $v$ are related to the timelike and spacelike coordinates
by the relations $u=(t-x)/2$ and $v=(t+x)/2$.

Relying on the covariance properties of the solutions we may choose a
standard representative $e_b^\mu$ in the class of spin frames inducing
$g$. Given the representative, a generic spin frame is of the form
$e'{}_a^\mu=e_b^\mu A^b_a$, where $A^b_a$ is an orthogonal matrix
$A^b_a\in\SO(1,1)$. The choice of a particular representative is a sort of
gauge fixing that simplifies the field equations. A convenient choice is
\be
\Vert e^a_\mu\Vert=\ep[\rho]=\left(\begin{array}{cc}
\rho & -\rho \\
\rho & \rho \\
\end{array}\right)\,,~~~~~~~\rho>0\,.
\label{SFrame}
\ee
The general solution of the Dirac equation (\ref{EDirac}) can be written in
the form (from now on we will consider $\alpha>0$; the generalization to
the case $\alpha<0$ is straightforward and will be omitted)
\be
\psi[\rho,\psi_u,\psi_v]={1\over(\alpha\rho)^{1/2}}
\left(\matrix{
\psi_v\cr\psi_u
}\right) \label{DiracSol}
\ee
where $\psi_u(u)$ and $\psi_v(v)$ are arbitrary functions of $u$ and $v$,
respectively.

Using equation (\ref{DiracSol}) in the field equations (\ref{EDilatone}) and
(\ref{EEinstein}) we obtain
\ba
&&\d_u\d_v(\ln \rho^2)-\rho^2{dU\over d\phi}=0\,,\label{eq-dil}\\
&&\d_u\d_v\phi-\rho^2 U(\phi)=0\,,\label{dynamical}\\
&&\rho^2\d_u\left({\d_u\phi\over \rho^2}\right)=F[\psi_u]
\,,\label{constr1}\\
&&\rho^2\d_v\left({\d_v\phi\over
\rho^2}\right)=G[\psi_v]
\,,\label{constr2}
\ea
where we have set
\ba
F[\psi_u]=i|\psi_u|^2\d_u
\left[\ln{\left({\psi_u^*\over\psi_u}\right)}\right]\,,\label{Fu}\\
G[\psi_v]=i|\psi_v|^2\d_v
\left[\ln{\left({\psi_v^*\over\psi_v}\right)}\right]\,.\label{Gv}
\ea

Let us remark that the gauge fixing (\ref{SFrame}) is not covariant, so it
destroys the explicit covariance of the field equations
(\ref{EDirac})-(\ref{EEinstein}). However, the covariance of equations
(\ref{EDirac})-(\ref{EEinstein}) can be restored {\it a posteriori} by
noticing that any solution of equations (\ref{eq-dil})-(\ref{constr2})
\be
(\rho(u,v),\psi_u(u),\psi_v(v),\phi(u,v))
\label{Sol}
\ee
identifies a whole class of physically equivalent solutions of equations
(\ref{EDirac})-(\ref{EEinstein}). Indeed, let us consider the
transformations induced by a vertical automorphism of $\Si$ on
$(\ep[\rho],\psi[\rho,\psi_u,\psi_v],\phi)$, where $\ep$ and $\psi$ are
defined in equations (\ref{SFrame}) and (\ref{DiracSol})
\be
(\ep,\psi,\phi)\mapsto
(\ep',\psi',\phi)\,.
\label{Aut}
\ee
Here $\ep'=\ell(S(u,v))\cdot\ep$, $\psi'=S(u,v)\cdot\psi$  and
\be
S(u,v)=\left(\begin{array}{cc}
e^{i\theta} & 0\\
0 & e^{-i\theta} \\
\end{array}\right)
\ee
is a generic matrix in $\Spin(1,1)$. It is straightforward to see that
$(\ep',\psi',\phi)$ is again a solution of equations
(\ref{EDirac})-(\ref{EEinstein}) corresponding to the same metric $g$
induced by $\ep$ (see equation (\ref{ind-met})).

An explicit invariance of equations (\ref{eq-dil})-(\ref{constr2}) is given by
the coordinate transformations preserving the conformal form of metric
(\ref{ansatz-metric})
\be
u=u(\tilde u)\,,~~~~~~~v=v(\tilde v)\,.\label{transf}
\ee
Under these transformations the conformal factor $\rho^2(u,v)$ of the
metric and the dilaton transform as (we consider without loss of
generality $(d\tilde u/du)>0$ and $(d\tilde v/ dv)>0$)
\ba
&&\rho^2(u,v)=\tilde \rho^2(\tilde u,\tilde v)\>{d\tilde u\over du}
\>{d\tilde v\over dv}\,,\label{tr-met}\\
&&\psi(u,v)= S(\tilde u,\tilde v)\tilde\psi_u(\tilde u,\tilde
v)\,,\label{tr-spin}\\
&&\phi(u,v)=\tilde\phi(\tilde u,\tilde v)\,,\label{tr-dilaton}
\ea
where $S$ is the two-dimensional $\Spin(1,1)$ matrix
\be
S(\tilde u,\tilde v)=\left(\matrix{\left[{d\tilde u\over
du}\right]^{-1/4}\left[{d\tilde v\over
dv}\right]^{1/4}&0\cr\cr
0&\left[{d\tilde u\over du}\right]^{1/4}\left[{d\tilde v\over
dv}\right]^{-1/4}\cr}\right)\,.\label{spin_matrix}
\ee
The field equations are explicitly conformally invariant under the
transformations (\ref{tr-met})-(\ref{tr-dilaton}). Hence, the rotation of
the spinor by the angle
\be
\theta=-{1\over 4}\ln\left({d\tilde u\over du}\right)+{1\over
4}\ln\left({d\tilde v\over dv}\right)\label{spin_angle}
\ee
preserves {\sl explicitly} the conformal invariance of the theory.
Finally, the functionals $F[\psi_u]$ and $G[\psi_v]$ transform as
\be
F[\psi_u]=\tilde F[\psi_{\tilde u}]\left({d\tilde
u\over du}\right)^2\,,~~~~~
G[\psi_v]=\tilde G[\psi_{\tilde v}]\left({d\tilde
v\over dv}\right)^2\,,\label{tr-fg}
\ee
under the transformation rule (\ref{transf}).

In the following sections we will find and discuss several exact solutions
for our model. The strategy is the following: starting from equations
(\ref{eq-dil})-(\ref{constr2}) we choose the functionals $F[\psi_u]$ and
$G[\psi_v]$ in equations (\ref{constr1})-(\ref{constr2}). (This is essentially
equivalent to choosing a particular class of spinors for the model.) Then we
solve the field equations (\ref{eq-dil})-(\ref{constr2}) for a given choice of
the potential $U(\phi)$. This completes the integration of the model.
\subsection{Solutions with constant-phase spinors\label{cph}}
The simplest family of exact solutions is obtained when the functionals
$F[\psi_u]$ and $G[\psi_v]$ are identically zero. In this case the metric
and the dilaton decouple from spinors and the system formally reduces to
pure DG. The most general class of spinors satisfying this condition is
given by
\be
\psi_u(u)=\psi_u^*(u)\,e^{-i\chi_u}\,,~~~~~~
\psi_v(v)=\psi_v^*(v)\,e^{-i\chi_v}\,,\label{const-phase}
\ee
where $\chi_u$ and $\chi_v$ are real constant parameters.

Two-dimensional pure DG has been discussed extensively in the literature.
So, without entering details, let us recall some interesting properties of
the model that will be useful later. It is well known that two-dimensional
pure DG can be completely integrated. A straightforward and powerful way
to prove this can be found in \cite{filippov}: the field equations
can be transformed into linear equations via a B{\"a}cklund-like
transformation and the general solution can be written as a function of
(two) free fields. A remarkable feature of this model is that the metric
tensor and the dilaton $\phi$ are actually depending only on one of the
two original free fields.  As a consequence, the general solution is
`static', i.e.\ the metric and the dilaton can be cast into the form
\cite{filippov}
\ba
&&ds^2=4f(u,v)dudv\,,\label{metric}\\
&&f(u,v)=h(\chi)\d_u\chi\,\d_v\chi\,,~~~~~\phi\equiv\phi(\chi)\,,
\label{mstatic}
\ea
where $\chi$ is a solution of the two-dimensional D'Alembert equation
$\d_u\d_v\chi=0$. (This property is nothing other that the generalized
Birkhoff theorem for this class of models.) Let us stress that our
definition of `staticity' is slightly different form the usual
definition that can be found in the literature (see, for instance, 
\cite{wald}). Here the concept of staticity refers to the fields, i.e.\ to
the metric and the dilaton, and not to the geometry. In this context
`staticity' simply means that a Killing vector, not necessarily timelike
and hypersurface orthogonal, does exist. Hence, for solutions of the form
(\ref{metric}) and (\ref{mstatic})  we can always choose a system of coordinates
such that the solution depends on a single coordinate, either the
spacelike or the timelike one. For instance, the two-dimensional
Schwarzschild metric $ds^2=(1-2M/x)dt^2-(1-2M/x)^{-1}dx^2$ is `static'
in the sense of (\ref{metric}) and (\ref{mstatic}) even though in the region
$x<2M$ the metric tensor depends on the timelike variable.

Let us make a short digression and discuss how the existence of solutions
of the form (\ref{metric}) and (\ref{mstatic}) is related to the structure of
the field equations. Let us consider equations (\ref{eq-dil})-(\ref{constr2}).
It is easy to prove the following theorem:
\begin{itemize}
\item{Theorem.} For any static solution (in the sense defined above)
there is a set of coordinates $(\tilde u,\tilde v)$ such that $\tilde
F[\psi_{\tilde u}]=\tilde G[\psi_{\tilde v}]=constant$. In the
coordinates $(\tilde u,\tilde v)$ the metric reads $ds^2=4h(\tilde
u+\tilde v)d\tilde u d\tilde v$.
\end{itemize}
The proof is very simple. Let us consider a static solution. The latter
can be cast into the form (\ref{metric}) and (\ref{mstatic}).  The general
solution of the D'Alembert equation in the lightcone coordinates can be
written as a sum of an arbitrary function of $u$ and an arbitrary function
of $v$. Setting $\chi=\tilde u(u)+\tilde v(v)$ and using the definition of
staticity (\ref{metric}) and (\ref{mstatic}) with $f=\rho^2$ it is
straightforward to verify that in the $(\tilde u, \tilde v)$ coordinates
equations (\ref{constr1}) and (\ref{constr2}) reduce to the form
\be
h\left({\phi'\over h}\right)'=
\tilde F[\psi_{\tilde u}]\,,~~~~~~
h\left({\phi'\over h}\right)'=\tilde G[\psi_{\tilde
v}]\,,\label{constr3}
\ee
where primes denote derivatives with respect to $\chi$. Finally, from equations
(\ref{constr3}) it follows immediately $\tilde F=\tilde G=constant$.

Hence, for static solutions we can always find a set of coordinates such that
the right-hand side of the two constraints (\ref{constr1}) and (\ref{constr2})
are identical. In this case the field equations (\ref{eq-dil}-\ref{constr2})
can be reduced to a system of ordinary differential equations. Since for static
solutions $F[\psi_u]=G[\psi_v]=constant$ is a necessary condition, the general
static solution of the model can be found by imposing directly
$F[\psi_u]=G[\psi_v]=constant$ in the field equations. Clearly, the above
condition is satisfied in the case of pure dilaton-gravity when $F$ and $G$ are
identically zero. In this case one can also prove the converse statement (see
\cite{filippov}), i.e.\ that the system possesses only static solutions. This
is achieved by the B{\"a}cklund-like transformation mentioned above. In the
case of $F\not=0$, $G\not=0$ the B{\"a}cklund-like transformation cannot be
implemented and non-static solutions exist.

In conclusion, the model with constant-phase spinors behaves essentially
like a pure DG system. The Hilbert tensor is identically zero and the
spinor field is decoupled both from the dilaton and the gravitational
field. Geometrically, for suitable choices of the dilaton potential
$U(\phi)$ black-hole-like solutions can be found. For instance, when
$U(\phi)\propto 1/\sqrt{\phi}$ the model can be interpreted as the
effective two-dimensional reduction of vacuum spherically symmetric
Einstein gravity. (The dilaton plays the role of the scale factor of the
sphere.) As a consequence, the model can be physically interpreted as a
black hole in a bath of massless non-interacting (ghost) spinors.
\subsection{Solutions with constant dilatonic potential\label{cpo}}
Since constant-phase spinors do not interact with either the dilaton or
with the gravitational field, let us look for a model in which the spinors
have a non-vanishing Hilbert tensor. Let us change our strategy and choose
a particular dilatonic potential, leaving undetermined the functionals
$F[\psi_u]$ and $G[\psi_v]$, i.e.\ the form of the spinor
(\ref{DiracSol}). The simplest choice is the constant dilaton potential
$U(\phi)=-\Lambda$. In this case the spacetime is flat and the system is
completely solvable.

The field equation for the dilaton (\ref{eq-dil}) and the dynamical
Einstein equation (\ref{dynamical}) read
\ba
&&\d_u\d_v(\ln\rho)=0\,,\label{eq-dil-lambda}\\
&&\d_u\d_v\phi+\Lambda\rho^2 =0\,.\label{dynamical-lambda}
\ea
Since equation (\ref{eq-dil-lambda}) does not involve the dilaton $\phi$ the
geometry of spacetime is decoupled both from the dilaton and the
spinorial field. So the first equation can be immediately solved and
the general solution can be written in the convenient form
\be
\rho^2(u,v)={d\tilde u\over du}{d\tilde v\over dv}\,,\label{flat-metric}
\ee
where $\tilde u$ and $\tilde v$ are arbitrary functions of $u$ and
$v$, respectively. From equation (\ref{flat-metric}) it is easy to see that
$\tilde u$ and $\tilde v$ are the lightcone coordinates of the
two-dimensional Minkowski spacetime. In the lightcone coordinates the
remaining equations become
\ba
&&\d_{\tilde u}\d_{\tilde v}\phi+\Lambda =0\,,\label{dynamical-lambda2}\\
&&\d^2_{\tilde u}\phi=\tilde F[\psi_{\tilde u}]\,,\label{constr1-lambda}\\
&&\d^2_{\tilde v}\phi=\tilde G[\psi_{\tilde v}]\,,\label{constr2-lambda}
\ea
where $\tilde F$ and $\tilde G$ are related to $F$ and $G$ in the
original $(u,v)$ variables by equations (\ref{tr-fg}). The metric
and the spinor read
\be
ds^2=4d\tilde u d\tilde v\,,~~~~~~
\psi(\tilde u, \tilde v)={1\over\sqrt{\alpha}}
\left(\matrix{\psi_{\tilde v}\cr
\psi_{\tilde u}}\right)\,.\label{flat}
\ee
(This corresponds to set directly $\rho=1$ in the field equations and use
flat coordinates; the functions $\psi_{\tilde u}$ and $\psi_{\tilde v}$
define the spinor in the Minkowskian system of coordinates.)

Equations (\ref{dynamical-lambda2})-(\ref{constr2-lambda}) can be easily
solved. The general solution for arbitrary $\tilde F$ and $\tilde G$ is
\be
\phi(\tilde u, \tilde v)=\phi_0+\phi_{\tilde u}\tilde u+\phi_{\tilde
v}\tilde v-\Lambda \tilde u\tilde v+U(\tilde u)+V(\tilde
v)\,,\label{gen-lambda}
\ee
where $\phi_0$, $\phi_{\tilde u}$, and $\phi_{\tilde
v}$ are constants, and
\be
U(\tilde u)=\int_{\tilde u_1}^{\tilde u}d\tilde u'\int_{\tilde
u_2}^{\tilde u'}d\tilde u''\tilde F[\psi_{\tilde u''}]\,,~~~~~
V(\tilde v)=\int_{\tilde v_1}^{\tilde v}d\tilde v'\int_{\tilde
v_2}^{\tilde v'}d\tilde v''\tilde G[\psi_{\tilde v''}]\,.
\ee

Even though the general solution is known, it is interesting to derive the
static solutions directly from the field equations. From the theorem of
the previous section it follows that for static solutions there exists a
set of coordinates $(\bar u,\bar v)$ such that
\be
F[\psi_{\bar u}]=G[\psi_{\bar v}]=\epsilon\,,\label{cond-static}
\ee
where $\epsilon$ is a real constant. In this system of coordinates the
fields can be chosen to depend only on $x=\bar v-\bar u$. (Alternatively,
on $t=\bar v+\bar u$.) So it is easy to integrate the field equations.  We
have two families of distinct solutions:
\begin{itemize}
\item{Solution (\ref{cpo}a):}
\ba
&&ds^2=4\rho_0^2 e^{k(\bar v\mp\bar u)}d\bar u d\bar v\,,\label{f-stat-a}\\
&&\phi=\phi_0-{\epsilon\over k}(\bar v\mp\bar u)\pm{\Lambda \rho_0^2\over
k^2}e^{k(\bar v\mp\bar u)}\,,~~~~~k\not=0\,,\label{phi-stat-a}\\
&&\psi={1\over(\alpha\rho_0)^{1/2}}e^{-k(\bar v\mp\bar
u)/4}\left(\matrix{\psi_{\bar
v}\cr \psi_{\bar u}}\right)\,;\label{spin-stat-a}
\ea
\item{Solution (\ref{cpo}b) (degenerate case, $\Lambda\not=0$,
$\epsilon\not=0$):}
\ba
&&ds^2=\pm 4{\epsilon\over\Lambda}d\bar u d\bar v\,,\label{f-stat-b}\\
&&\phi=\phi_0+\phi_1(\bar v\mp\bar u)+{\epsilon\over
2}(\bar v\mp\bar u)^2\,,\label{phi-stat-b}\\
&&\psi={[\pm(\Lambda/\epsilon)]^{1/4}\over\sqrt{\alpha}}
\left(\matrix{\psi_{\bar
v}\cr \psi_{\bar u}}\right)\,;\label{spin-stat-b}
\ea
\end{itemize}
where $\rho_0>0$, $k$, $\phi_0$, and $\phi_1$ are integration constants.
Note that both solutions are describing a flat two-dimensional manifold,
as expected for a constant dilaton potential (see equation
(\ref{EDilatone})). However, the two solutions correspond to different
choices of the spinor.  This can be easily seen using Minkowskian
coordinates and comparing the two solutions. In the Minkowskian
coordinates $(\tilde u,\tilde v)$ the dilaton and the spinor are
\begin{itemize}
\item{Solution (\ref{cpo}a):}
\ba
&&\phi=\tilde\phi_0-\Lambda\tilde u
\tilde v-\tilde\epsilon\ln{|\tilde u\tilde
v|}\,,\label{phi-stat-a2}\\
&&\psi={1\over\sqrt{\alpha}}\left(\matrix{\psi_{\tilde v}\cr
\psi_{\tilde u}}\right)\,,\label{spin-stat-a2}
\ea
where $\tilde\phi_0$ and $\tilde\epsilon=\epsilon/k^2$ are constants and
$\psi_{\tilde u}$, $\psi_{\tilde v}$ are defined by the relations
\be
\tilde F[\psi_{\tilde u}]={\tilde\epsilon\over\tilde u^2}\,,~~~~~~
\tilde G[\psi_{\tilde v}]={\tilde\epsilon\over\tilde v^2}\,;
\label{FG-stat-a}
\ee
\item{Solution (\ref{cpo}b):}
\ba
&&\phi=\phi_0+\tilde\phi_1(\tilde v\mp\tilde u)\pm{\Lambda\over
2}(\tilde v\mp\tilde u)^2\,,\label{phi-stat-b2}\\
&&\psi={1\over\sqrt{\alpha}}
\left(\matrix{\psi_{\tilde v}\cr
\psi_{\tilde u}}\right)\equiv {[\pm(\Lambda/\epsilon)]^{1/4}
\over\sqrt{\alpha}}\left(\matrix{\psi_{\bar v}\cr
\psi_{\bar u}}\right)\,,
\ea
where $\tilde\phi_1$ is a constant and
\be
\tilde F[\psi_{\tilde u}]=\tilde
G[\psi_{\tilde v}]=\pm\Lambda\,.\label{FG-stat-b}
\ee
\end{itemize}
In the case (\ref{cpo}b) the spinor is chosen so that the two functionals
$\tilde F$ and $\tilde G$ are equal to $\pm\Lambda$ in the flat system of
coordinates; in the first case the spinors are chosen so that the
functionals $\tilde F$ and $\tilde G$ are not constant in the Minkowskian
coordinates. Note that the solution (\ref{cpo}b) is singular for
$\epsilon=0$ and has no counterpart in the pure DG theory unless
$\Lambda=0$. (This is evident in the Minkowskian coordinates because
$\tilde F$ and $\tilde G$ cannot be zero for $\Lambda\not=0$, see equations
(\ref{FG-stat-b}).) Solution (\ref{cpo}a) reduces to the pure DG solution
when $\tilde\epsilon=0$. Finally, both solutions can be obtained directly
from (\ref{gen-lambda}) using (\ref{FG-stat-a}) and (\ref{FG-stat-b}).

To conclude the discussion of the solutions with constant dilatonic
potential let us find a representation for the spinor. Let us consider
solution (\ref{cpo}b). Since the metric is flat plane-wave spinors
are the natural choice. In the $(\tilde u,\tilde v)$ coordinates
plane-wave spinors are
\be
\psi={1\over 2\sqrt{2\pi}}\left(\matrix{e^{ip\tilde v}\cr
e^{ip\tilde u}}\right)\,,\label{plane-wave-spin}
\ee
where $p\alpha/(4\pi)=\pm\Lambda$. Now, let us consider solution
(\ref{cpo}a). In this case plane-wave spinors (\ref{plane-wave-spin})
are not compatible with equations (\ref{phi-stat-a2}) and (\ref{spin-stat-a2}) 
and equation (\ref{FG-stat-a}) because the solution is singular for $\tilde u=0$
and $\tilde v=0$. We can choose instead
\be
\psi={1\over 2\sqrt{2\pi}}\left(\matrix{|\tilde
v|^{\pm ip-1/2}\cr\cr
|\tilde u|^{\pm ip-1/2}}\right)\,,\label{plane-wave-spin-a}
\ee
where $p\alpha/(4\pi)=\tilde\epsilon$ and signs correspond to $\tilde
v>0$, $\tilde v<0$ and $\tilde u>0$, $\tilde u<0$, respectively.  Spinors
in equation (\ref{plane-wave-spin-a}) satisfy the conditions
(\ref{FG-stat-a}) and correspond (apart from a constant phase factor) to
plane waves in the $(\bar u,\bar v)$ system of coordinates.
\subsection{Solutions with linear dilatonic potential\label{lpo}}
We have seen that for a constant dilatonic potential the model is
completely solvable. However, the gravitational degree of freedom
represented by the conformal factor $\rho$ decouples dynamically both from
the dilaton and the spinor field. As a consequence, the two-dimensional
spacetime is simply the Minkowski spacetime and the model can be
interpreted as a bath of interacting spinor and dilaton fields in a flat
background. This is not surprising. Indeed, the curvature of the spacetime
is proportional to the derivative of the dilatonic potential, as one
can easily verify from the field equation (\ref{EDilatone}).

So it is interesting to explore models with curved geometry and see how
the presence of spinors modifies the structure of the spacetime. The
simplest non-flat model is identified by the linear dilatonic potential
$U(\phi)=-\lambda\phi$. (In pure DG the model is known as `pure
string-inspired dilaton-gravity' or the `Jackiw-Teitelboim model' and has
been investigated extensively in the literature; see, for instance, 
\cite{2d1}-\cite{2d5} and references quoted therein.) Even though the
dynamics of the spacetime remains decoupled from the dilaton and the
spinor field (the manifold has constant curvature $\lambda$) the structure
of the model is richer with respect to the flat case. For instance, pure
string-inspired two-dimensional DG can be interpreted as the effective
reduced theory of vacuum axisymmetric gravity plus cosmological constant
in $2+1$ dimensions (see, for example, \cite{ap}). It is well known
that axisymmetric black holes of constant curvature do exist in $2+1$
dimensions \cite{BTZ}. So the model with linear dilatonic potential can be
interpreted as describing a bath of spinors in a black hole background and
is worthwile investigating.

Let us consider a linear dilatonic potential and restrict attention on
static solutions. Solving the field equations we find three families of
solutions (we consider for simplicity the fields depending on $x$.
Static solutions depending on $t$ can be obtained by the substitution
$x\to t$, $\lambda\to -\lambda$)
\begin{itemize}
\item{Solution (\ref{lpo}a):}
\ba
&&\rho^2(x)=\sigma{2\over\lambda a^2}{1\over P(x)^2}\,,\\
&&\phi(x)=\phi_0 Q(x)+{\epsilon a^2\over
2}\left[{x\over a}Q(x)-1\right]\,,
\ea
where
\ba
&&P(x)=\sinh{\left({x-x_0\over
a}\right)}\,,~~~Q(x)=\coth{\left({x-x_0\over a}\right)}\,,~~~{\rm for}~~
\sigma=1\,,\label{a1}\\
&&P(x)=\cosh{\left({x-x_0\over
a}\right)}\,,~~~Q(x)=\tanh{\left({x-x_0\over a}\right)}\,,~~~{\rm for}~~
\sigma=-1\,.\label{a2}
\ea
\item{Solution (\ref{lpo}b):}
\ba
&&\rho^2(x)={2\over\lambda a^2}{1\over\sin^2{\left({x-x_0\over
a}\right)}}\,,\\
&&\phi(x)=\phi_0\cot{\left({x-x_0\over a}\right)}-{\epsilon a^2\over
2}\left[{x\over a}\cot{\left({x-x_0\over a}\right)}-1\right]\,.
\ea
\item{Solution (\ref{lpo}c):}
\ba
&&\rho^2(x)={2\over\lambda (x-x_0)^2}\,,\\
&&\phi(x)={\phi_0\over x-x_0}+{\epsilon\over 6}(x-x_0)^2\,;
\ea
where $x_0$, $\phi_0$ and $a\not=0$ are constants of integration and
$\epsilon$ is defined as in the previous section.
\end{itemize}

It is straightforward to verify that the three solutions describe a
two-dimensional spacetime with constant Ricci scalar $R=-\lambda$, as
expected from the dilaton field equation. However, the presence of matter
fields, i.e.\ the dilaton and the spinor, breaks the isometry
group of the de Sitter metric \cite{ks}. As a consequence, the three
solutions are diffeomorphism inequivalent solutions of the field
equations and correspond to different physical descriptions. Let us
see this in detail and define the spacelike coordinate
\be
y=-\int \rho^2(x) dx\,.
\ee
In the $(t,y)$ coordinates the line element has the standard
`Schwarz\-schild\--li\-ke' form
\be
ds^2=f(y)dt^2-f(y)^{-1}dy^2\,,~~~~~~ f(y)={\lambda\over
2}\left[y^2+\gamma\right]\,,
\ee
where
\ba
&&y={2\over\lambda a}\coth{x-x_0\over a}\,,~~~~\gamma=-{4\over\lambda^2
a^2}\,,\\
&&y={2\over\lambda a}\tanh{x-x_0\over a}\,,~~~~\gamma=-{4\over\lambda^2
a^2}\,,
\ea
for solutions (\ref{lpo}a) and
\ba
&&y={2\over\lambda a}\cot{x-x_0\over a}\,,~~~~\gamma={4\over\lambda^2
a^2}\,,\\
&&y={2\over\lambda (x-x_0)}\,,~~~~\gamma=0\,,
\ea
for solutions (\ref{lpo}b) and (\ref{lpo}c), respectively. The Penrose
diagrams for the above solutions can be found in \cite{ks}. The
physical meaning of the solutions (\ref{lpo}a)-(\ref{lpo}c) is
obvious: the three solutions describe the dilaton and the spinor field
in different regions of the de Sitter spacetime separated by Killing
horizons.

Solutions (\ref{lpo}a) can be cast in a simpler and unique form using
reparametrizations in $u$ and $v$. Setting for simplicity $x_0=0$ we have
\ba
&&ds^2={8\over\lambda}{d\tilde u d\tilde v\over (\tilde v-\tilde u)^2}\,,\\
&&\phi=\phi_0{\tilde v+\tilde u\over \tilde v-\tilde u}+\tilde\epsilon
\left[{\tilde v+\tilde u\over \tilde v-\tilde u}
\ln{\left|{\tilde v\over\tilde u}\right|}-2\right]\,,\label{sol-a-2}\\
&&\psi=\left|{\lambda\over 2}\right|^{1/4}\sqrt{|\tilde v-\tilde
u|\over\alpha}\left(\matrix{\psi_{\tilde v}
\cr\psi_{\tilde u}}\right)\,,
\ea
where
\be
u={a\over 2}\ln|\tilde u|\,,~~~~v={a\over 2}\ln|\tilde v|\,,
\ee
and $\tilde\epsilon=\epsilon a^2/4$. The two choices (\ref{a1}) and
(\ref{a2}) correspond to the regions $\tilde u\tilde v>0$ and $\tilde u
\tilde v<0$, respectively. In the $(\tilde u,\tilde v)$ coordinates
plane-wave spinors read
\be
\psi=\sqrt{{|\tilde v-\tilde u|\over 8\pi}\sqrt{\left|{\lambda\over
2}\right|}}\left(\matrix{|\tilde v|^{\pm ip-1/2}\cr\cr
|\tilde u|^{\pm ip-1/2}}\right)\,,
\ee
where $\tilde\epsilon$ is defined as a function of $p$ as below equation
(\ref{plane-wave-spin-a}) and signs correspond to $\tilde v>0$,
$\tilde v<0$ and $\tilde u>0$, $\tilde u<0$, respectively.

In conclusion, the linear-potential model describes a spinor field
interacting with the dilaton in a fixed background with constant
curvature. The geometry of spacetime is dynamically decoupled both from
the spinor and the dilaton. This happens both for solutions with a Killing
vector and for general solutions because the curvature of spacetime is
completely determined by the dilatonic potential. In the next section we
will discuss a more interesting case:  the exponential dilatonic
potential.
\subsection{Solutions with exponential dilatonic potential\label{epo}}
Let us consider a potential of the form
\be
U(\phi)=\lambda e^{k\phi}\,,
\ee
where $\lambda\not=0$ and $k\not=0$ are constants. The field equations can
be solved exactly in the static sector. We find four distinct families of
solutions (again we consider the fields depending on $x$):
\begin{itemize}
\item{Solution (\ref{epo}a):}
\ba
&&\rho^2(x)={\gamma^2\over k\lambda}{e^{-k(\phi_0+ax)}
\over\cosh{\left[\gamma(x-x_0)\right]}}\,,\\
&&\phi(x)=\phi_0+ax-{1\over
k}\ln{\left\{\cosh{\left[\gamma(x-x_0)\right]}\right\}}\,,\\
&&\epsilon=ka^2-\gamma^2/k\,;
\ea
\item{Solution (\ref{epo}b):}
\ba
&&\rho^2(x)=-{\gamma^2\over k\lambda}{e^{-k(\phi_0+ax)}
\over\sinh{\left[\gamma(x-x_0)\right]}}\,,\\
&&\phi(x)=\phi_0+ax-{1\over
k}\ln{\left\{\sinh{\left[\gamma(x-x_0)\right]}\right\}}\,,\\
&&\epsilon=ka^2-\gamma^2/k\,;
\ea
\item{Solution (\ref{epo}c):}
\ba
&&\rho^2(x)=-{\gamma^2\over k\lambda}{e^{-k(\phi_0+ax)}
\over\sin{\left[\gamma(x-x_0)\right]}}\,,\\
&&\phi(x)=\phi_0+ax-{1\over
k}\ln{\left\{\sin{\left[\gamma(x-x_0)\right]}\right\}}\,,\\
&&\epsilon=ka^2+\gamma^2/k\,;
\ea
\item{Solution (\ref{epo}d):}
\ba
&&\rho^2(x)=-{1\over k\lambda}{e^{-k\sqrt{\epsilon\over
k}(x-x_0)}\over x-x_0}\,,\\
&&\phi(x)=\sqrt{\epsilon\over k}(x-x_0)-{1\over
k}\ln{(x-x_0)}\,;
\ea
where $a$, $\gamma\not=0$, and $x_0$ are constants of integration.
\end{itemize}

It is straightforward to prove that the four families of solutions are
actually distinct. This can be done by calculating the Ricci scalar. From
the dilaton field equation (\ref{EDilatone}) we have
\be
R={\lambda\over k}e^{k\phi}\,.\label{ricci-exp1}
\ee
So the Ricci scalar is proportional to the inverse of
$\cosh(\gamma(x-x_0))$, $\sinh(\gamma(x-x_0))$,
$\sin(\gamma(x-x_0))$ and $(x-x_0)$ for the four solutions, respectively.
In the first
case the manifold is regular for any value of the spacelike coordinate
$x$. Conversely, for the solutions (\ref{epo}b)-(\ref{epo}d)  the
manifolds have singular points: $x=x_0$ for solutions (\ref{epo}b)  and
(\ref{epo}d), and $x=x_0=n\pi/\gamma$ for the case (\ref{epo}c), where
the dilaton field blows up. The discussion of the geometrical structure
and of the global properties of these solutions is beyond the purpose of
this paper and will be considered elsewhere. However, let us spend a few
words on some particular cases. Solutions (\ref{epo}c) have no limit
for $\epsilon=0$, i.e.\ they are not solutions of the pure DG theory
without spinors. Conversely, solutions (\ref{epo}a), (\ref{epo}b), and
(\ref{epo}d) admit a pure DG limit. Solutions (\ref{epo}a) and
(\ref{epo}b) can be rewritten in a more interesting form using
reparametrizations in $u$ and $v$. Using the coordinates
\be
u={1\over\gamma}\ln|\bar u|\,,~~~~~v={1\over\gamma}\ln|\bar v|\,,
\ee
solutions (\ref{epo}a) and (\ref{epo}b) read
\begin{itemize}
\item{Solution (\ref{epo}a):}
\ba
&&ds^2=\pm{8\over\lambda k}e^{-k\phi_0}\left|{\bar u\over\bar
v}\right|^{\kappa}{d\bar u d\bar v\over \bar v^2+\bar u^2}\,,\\
&&\phi=\phi_0+{1\over k}\ln{\left[\left|{\bar v\over\bar
u}\right|^{\kappa}{2|\bar u\bar v|\over \bar v^2+\bar u^2}\right]}\,.
\ea
\item{Solution (\ref{epo}b):}
\ba
&&ds^2=\mp{8\over\lambda k}e^{-k\phi_0}\left|{\bar u\over\bar
v}\right|^{\kappa}{d\bar u d\bar v\over \bar v^2-\bar u^2}\,,\\
&&\phi=\phi_0+{1\over k}\ln{\left[\left|{\bar v\over\bar
u}\right|^{\kappa}{2|\bar u\bar v|\over \bar v^2-\bar u^2}\right]}\,;
\ea
\end{itemize}
where the upper signs refer to the I and IV quadrants of the $(\bar
u,\bar v)$ plane, the lower signs refer to the II and III quadrants, and
$\kappa=\pm\sqrt{1+\epsilon k/\gamma^2}$. Note that $\kappa=\pm 1$ for
the pure DG case. The Ricci scalar in the new coordinates is (the signs
refer to solutions (\ref{epo}a) and (\ref{epo}b), respectively):
\be
R=\lambda k e^{k\phi_0}\left|{\bar v\over\bar
u}\right|^{\kappa}{2|\bar u\bar v|\over \bar v^2\pm\bar
u^2}\,.\label{ricci-exp2}
\ee
From equation (\ref{ricci-exp2}) we see that the geometrical structure of the
solutions depends on the value of the parameters $\kappa$ and $k$. First
of all, the sign of the curvature is constant for fixed $\lambda$ and $k$,
as expected from equation (\ref{ricci-exp1}). Furthermore, when $\kappa=\pm
1$, i.e.\ when there are no spinors, the spacetime (\ref{epo}a) is never
singular and the spacetime (\ref{epo}b) is singular only for $\tilde
v^2-\tilde u^2=0$. This picture changes drastically if $\kappa\not=1$. In
this case the spacetime is singular on $\bar u=0$ ($\kappa>1$) and $\bar
v=0$ ($\kappa<-1$). When spinors are present the dilaton field blows up
for $\bar u=0$ and $\bar v=0$. In pure DG this happens only for $\bar u=0$
or $\bar v=0$, depending on the $\kappa$-branch chosen.
\section{Conclusions and perspectives}
In this paper we have presented and discussed some exact solutions of
two-dimensional DG coupled to massless spinors. In our formalism the
gravitational field is described by new variables, the so-called spin
frames, related to spin structures on the spacetime manifold and defined
without any reference to any preferred background metric. The theory is
geometrically well defined on spin manifolds and describes gravity
dynamically coupled to spinorial fields. Hence, the formalism describes
both the effect of gravity on spinors and the effect of spinors on the
gravitational field for a large class of physical manifolds.

We have focused attention on three physically interesting choices of the
dilatonic potential: constant, linear and exponential potentials. In the
first case the theory is completely solvable and represents a remarkable
new example in the family of two-dimensional DMG integrable systems. We
have integrated locally the field equations and discussed briefly the
properties of the solutions. Even though in this simple model the
spacetime is flat and there are no black hole solutions, several
interesting conclusions can be drawn.  Indeed, the presence of the spinor
field changes dramatically the structure of the solutions with respect to the 
pure DG case. For instance, the Birkhoff theorem is no longer valid and
solutions without any symmetries appear. This important property seems
not to depend on the particular model chosen, i.e.\ on the dimension of
the spacetime and on the form of the dilatonic potential. Since
two-dimensional DG theories can be interpreted, for suitable choices of
the dilatonic potential, as effective dimensionally reduced models of
gravity in $N>2$ dimensions, the presence of spinors may have interesting
consequences on the physics of black holes in higher dimensions. So the
investigation of effective models describing three- or four-dimensional
black holes interacting with the spinor field is worthwile 
investigating and may constitute a fruitful subject for future research.  A
small step in this direction has been done in section \ref{lpo} where we
have discussed the linear dilatonic potential. Indeed, the
Jackiw-Teitelboim model can be interpreted as the effective reduced theory
of vacuum axisymmetric gravity plus cosmological constant in $2+1$
dimensions. Since axisymmetric black holes of constant curvature do exist
in $2+1$ dimensions \cite{BTZ}, the model describes a bath of spinors in a
black hole background.

Furthermore, the Hamiltonian formulation and quantization of the theory
are subjects worth to be explored. In particular, the quantization of pure
DG with constant dilatonic potential has been studied in detail in the
literature \cite{2d1}-\cite{2d5}. (The main results of these
investigations are summarized in the report \cite{jacksm}.) Since the
presence of spinors does not destroy the complete solvability of the
system, the quantization of SDG with constant potential might not be very
different from the quantization of the CGHS model \cite{CGHS} and the
well known techniques developed for the CGHS model might be successfully
used for SDG.

Finally, let us mention that an alternative viewpoint on topological
two-dimensional gravity has been recently suggested by one of the authors
(in collaboration with Volovich, see \cite{22,23}) on the basis
of a metric-affine formalism (or first-order formalism \`a la
Palatini). It is well known that in two dimensions the metric and the
metric-affine formalisms are not equivalent (see also 
\cite{24,25}). In the future we aim to discuss the application of spin
frames to this alternative formalism.

\null\medskip
\noindent
{\large\bf Acknowledgments}

\noindent
One of the authors (MC) is very grateful to Vittorio de Alfaro,
Alexandre T Filippov, Jorma Louko and Germar Schr\"{o}der for
interesting discussions and useful suggestions on various questions
connected to the subject of this paper. MC is supported by a Human
Capital and Mobility grant of the European Union, contract number
ERBFMRX-CT96-0012.

\appendix{The Formalism of Spin Frames}
In this appendix, we briefly outline the spin frame formalism (see also
\cite{VB,Spinors}) and set up notations.

Let us consider a two-dimensional, connected, orientable manifold $M$ that
allows metrics of signature $(1,1)$. Since $\dim(M)=N=2$, the relevant
groups are $O(1)\simeq Z_2$, $\SO(1,1)\simeq \Re$, and
$\Spin(1,1)\simeq\Re$ which is embedded into $\calC^+(1,1)\simeq\Re^2$ as
a branch of hyperbola.  Due to the choice $\dim(M)=N=2$, the group
covering $\ell:\Spin(1,1)\arr\SO(1,1)$ is trivial and one-to-one when we
restrict ourselves to the connected components of unity, as we have always
understood. (For the general formalism in $N>2$ dimensions, see e.g.\
\cite{Spinors}.) 

This topological requirement ensures that we may consider reductions of
the tangent bundle $TM$ to the structure group $ O(1)\times O(1)\subset
\GL(2)$ or, equivalently, that there exists a splitting sequence of the
form
\be
0  \rarr
T  \op\rharr^{i_+}
TM \op{\buildrel\lharr \over\rarr}_{p_-}^{i_-}
S  \rarr
0\,,
\ee
where $T$ and $S$ are two line bundles over $M$.

The manifold $M$ is interpreted as the spacetime. One should also require
$M$ to be a spin manifold, i.e.\ to have vanishing second Stiefel-Whitney
class $w_2(M,Z_2)=0$. This is again a topological condition on $M$. Of
course, since in two dimensions any orientable manifold is a spin
manifold, we do not impose any further restriction on $M$.

Let us consider a $\Spin(1,1)$-principal bundle $\Si$ over $M$, i.e.
a principal $\Re$-bundle. A spin frame on $\Si$ is a global vertical
principal morphism $\La:\Si\arr L(M)$ of $\Si$ in the frame bundle $L(M)$;
namely, the following diagrams commute:
\be
\begin{array}{ccl}
 \Si  & \op\longrightarrow^{\La} &  L(M) \\
  \put(0,5){\vector(0,-1){20}} &
\put(10,5){\vector(-1,-1){20}}
& \\
 M &  &
\end{array}\qquad\qquad
\begin{array}{rcc}
 \>\Si & \op\longrightarrow^{\La} &  L(M) \\
 \put(-20,-5){${}_{R_S}$}\put(-5,5){\vector(0,-1){20}} &
& \put(5,5){\vector(0,-1){20}}\put(10,-5){${}_{R_{\hat\ell(S)}}$}\\
\>\Si & \op\longrightarrow^{\La} &  L(M)
\end{array}
\ee
where $R$ denotes the canonical right action defined on any principal bundles.
Here $\hat\ell=i\circ\ell$, $i:\SO(1,1)\arr\GL(2)$ is the canonical
immersion and $\ell:\Spin(1,1)\arr\SO(1,1)$ is the covering map.

A spin principal bundle $\Si$ is called structure bundle for $M$ if
at least one spin frame on $\Si$ does exist. Of course, there are spin
principal bundles that are not structure bundles for $M$; however, since
$M$ is a spin manifold there is at least one structure bundle for $M$.

Let us consider now a structure bundle $\Si$ for $M$ and a spin frame
$\La$ on it. A $\SO(1,1)$-principal subbundle $\Im(\La)$ is induced in
$L(M)$ and, according to \cite{Spinors}, uniquely identifies a
metric $g(\La)$ of signature $(1,1)$ called the metric induced by
$\La$ (see below equation (\ref{ind-met})). Let us remark that $\La$ is onto
the special orthonormal frame bundle $\SO(M,g(\La))\equiv\Im(\La)$ and is
a covering. Thus $(\Si,\La)$ is a spin structure on $(M,g(\La))$ in the
sense of \cite{Mil,GP}.  Conversely, if $(\Si,\tilde\La)$ is a spin
structure on $(M,g)$, by composing $\tilde\La:\Si\arr\SO(M,g)$ with the
canonical immersion $i_g:\SO(M,g)\arr L(M)$ we obtain
$\La_g=i_g\circ\tilde\La$, which is a spin frame on $\Si$. 

Since spin frames induce a metric uniquely (but not vice versa) they
represent natural candidates as dynamical fields to describe gravity.
However, in order to be interpreted as fields, they should have some
value at any point of $M$. In other words, spin frames on $\Si$ must be
sections of some bundle over $M$. This can be easily achieved by
considering the associated bundle $\Si_\rho=(\Si\times_M
L(M))\times_\rho\GL(2)$ through the action
\be
\rho:\Spin(1,1)\times\GL(2)\times\GL(2)\arr\GL(2):(S,J,e)\mapsto J\cdot
e\cdot \ell(S^{-1})\,.
\ee
Finally, one may also show that there is a one-to-one correspondence
between spin frames on $\Si$ and sections of the spin frame bundle
$\Si_\rho$.

Points in $\Si_\rho$ are represented by $[\sig,\del,e_a^\mu]_\rho$, where
$[\cdot]_\rho$ denotes the orbit under the action, $\sig\in\Si$, $\del\in
L(M)$, and $e_a^\mu\in\GL(2)$; thus $(x^\mu,e_a^\mu)$ are local
coordinates on $\Si_\rho$. Since there is no preferred choice of a unit
point, $\Si_\rho$ is neither a principal bundle nor a group bundle, even
though it has fibres that are diffeomorphic to the group $\GL(2)$.

The construction of the induced metric presented above provides a
canonical global epimorphism $g_{_\Si}:\Si_\rho\arr\Met(M;1,1)$, where
$\Met(M;1,1)$ is the bundle of all metrics of signature $(1,1)$ on
$M$;  this morphism is called inducing metric morphism and is
locally given by
\be
g_{_\Si}(x^\mu,e_a^\mu)=(x^\mu,g_{\mu\nu})\,,
\qquad\qquad
g_{\mu\nu}=e^a_\mu\>\eta_{ab}\>e^b_\nu\,,
\label{ind-met}\ee
where $e^a_\mu$ is the inverse matrix of $e_a^\mu$ and $\eta_{ab}$ is the
canonical diagonal matrix of signature $(1,1)$. Latin indices are lowered
and raised by $\eta_{ab}$ and Greek indices by the induced metric
$g_{\mu\nu}$; thus one can write simply $e_{a\mu}$ regardless if the
latter is given by $\eta_{ab}\>e^b_\mu$ or $e_a^\nu\> g_{\mu\nu}$ since
these two expressions coincide.

Of course, metric actions written on $\Met(M;1,1)$ can be canonically
pulled-back on $\Si_\rho$ along the inducing metric morphism. If the
metric Lagrangian is generally covariant then the pull-back Lagrangian on
$\Si_\rho$ is covariant with respect to the group $\Aut(\Si)$ of any
(not necessarily vertical) automorphism of $\Si$. This group acts
on $\Si_\rho$ as
\be
\Phi\in\Aut(\Si)
\quad\leadsto\quad
\Phi_\rho:\Si_\rho\arr\Si_\rho:[\sig,\del,e_a^\mu]_\rho\mapsto
[\Phi(\sig),L_f(\del),e_a^\mu]_\rho\,,
\ee
where $L_f:L(M)\arr L(M)$ is the natural lift to the frame bundle of the
diffeomorphism $f:M\arr M$ induced on $M$ by $\Phi$. Locally we have
\be
\begin{array}{lcl}
\Phi(x,S)=(f(x),\vp(x)\cdot S)
&\leadsto
&\Phi_\rho(x,e_a^\mu)=(f(x), J^\mu_\nu(x)\>
e^\nu_b\>\ell^b_a(\vp^{-1}(x)))\,,
\\
&&\\
\vp:U\subset M\arr\Spin(1,1)\,,
&&J^\mu_\nu(x)={\del f^\mu\over\del x^\nu}(x)\,.
\end{array}
\label{Spin-frames-transf}\ee

Equation (\ref{Spin-frames-transf}) shows that spin frames behave in a
very different way from standard frames with respect to symmetries (i.e.\
transformation laws). Indeed, the transformation laws of a frame with
respect to a diffeomorphism $f:M\arr M$ is
\be
e'{}_a^\mu(x')= J^\mu_\nu e_a^\nu(x)\,,
\qquad
x'=f(x),\>\> J^\mu_\nu=\del_\nu f^\mu\,.
\ee
These two different behaviours account for different Lie derivatives used
in literature and then for different conserved quantities.  Furthermore,
they also account for different ways of gluing patches together, so they
may be relevant for exact solutions too.

In order to consider interactions with spinors we choose a
$\Spin(1,1)$-representation $\la$ on a suitable vector space $V$. For
instance, in two dimensions we can choose $V=\Co^2$ and the
representation induced by the two-dimensional Dirac matrices
\be
\ga_0=\left(\begin{array}{rr}
0 & 1 \\
1 & 0 \\
\end{array} \right)\,,
\qquad
\ga_1=\left(\begin{array}{rr}
0 & -1 \\
1 & 0 \\
\end{array} \right)\,.
\ee
Let us consider the associated spinor bundle $\Si_\la=\Si\times_\la
V$ whose sections are, by definition, spinor fields on $M$. A point of
$\Si_\la$ is given by $[\sig, \psi]_\la$, where $[\cdot]_\la$ denotes the
orbit, $\sig\in\Si$, and $\psi\in V$; the local coordinates are
$(x^\mu,\psi^i)$. Again we have an action of $\Aut(\Si)$ on $\Si_\la$
given by
\be
\Phi_\la[\sig,\phi]_\la=[\Phi(\sig),\psi]_\la\,,
\qquad\qquad
\Phi_\la(x^\mu,\psi^i)=(f^\mu(x),\la^i_j(\vp(x))\psi^j)\,.
\ee
Finally, dilatons (scalar fields) are sections of the trivial bundle
$\De=M\times\Co$. On $\De$ we have a (trivial) action of $\Aut(\Si)$ given
by
\be
\Phi_\De(x,\phi)=(f(x),\phi)\,.
\ee

\end{document}